\documentclass[aps,prx,twocolumn,amsmath,amssymb,bibnotes,longbibliography]{revtex4-1}
\usepackage{textcomp}
\usepackage{pifont}
\usepackage{mathptmx}
\usepackage{graphicx}
\usepackage{enumerate}
\usepackage{titlesec}

\usepackage{hyperref}
\hypersetup{
pdftitle={Photon-Mediated Quantum Gate between Two Trapped Neutral Atoms in an Optical Cavity},
pdfauthor={Stephan Welte, Bastian Hacker, Severin Daiss, Stephan Ritter, and Gerhard Rempe},
colorlinks=true, linkcolor=[RGB]{45,48,146}, citecolor=[RGB]{45,48,146}, filecolor=[RGB]{45,48,146}, urlcolor=[RGB]{45,48,146}}

\newcommand{\unit}[1]{\ensuremath{\,\mathrm{#1}}}
\newcommand{\micro}{\ensuremath{\textnormal\textmu}}
\newcommand{\bra}[1]{\langle{#1}|} %bra
\newcommand{\ket}[1]{|{#1}\rangle} %ket
\newcommand{\up}{{\uparrow}} % arrow with spacing as symbol, not as delimiter
\newcommand{\down}{{\downarrow}} % arrow with spacing as symbol, not as delimiter
\newcommand{\0}{\hphantom{0}}\newcommand{\m}{\hphantom{\ensuremath{{-}}}}

\DeclareTextFontCommand{\emph}{\textit}

\makeatletter % make document compile without hyperref
\@ifpackageloaded{hyperref}{\newcommand{\positionlabel}[1]{\hypertarget{#1}{}}}{\newcommand{\positionlabel}[1]{}}
\@ifpackageloaded{hyperref}{}{}
\@ifpackageloaded{hyperref}{\newcommand{\figref}[1]{\hyperlink{#1}{\ref*{#1}}}}{\newcommand{\figref}[1]{\ref{#1}}}
\makeatother

\begin{document}

\title{Photon-Mediated Quantum Gate between Two Neutral Atoms in an Optical Cavity}

\author{Stephan~Welte}
\email[To whom correspondence should be addressed. Email: ]{stephan.welte@mpq.mpg.de}
\author{Bastian~Hacker}
\author{Severin~Daiss}
\author{Stephan~Ritter}
\thanks{Present address: TOPTICA Photonics AG, Lochhamer Schlag 19, 82166 Gr\"afelfing, Germany}
\author{Gerhard~Rempe}

\affiliation{Max-Planck-Institut f\"ur Quantenoptik, Hans-Kopfermann-Strasse 1, 85748 Garching, Germany}

\begin{abstract}
Quantum logic gates are fundamental building blocks of quantum computers. Their integration into quantum networks requires strong qubit coupling to network channels, as can be realized with neutral atoms and optical photons in cavity quantum electrodynamics. Here we demonstrate that the long-range interaction mediated by a flying photon performs a gate between two stationary atoms inside an optical cavity from which the photon is reflected. This single step executes the gate in $2\unit{\micro s}$. We show an entangling operation between the two atoms by generating a Bell state with $76(2)\%$ fidelity. The gate also operates as a CNOT. We demonstrate $74.1(1.6)\%$ overlap between the observed and the ideal gate output, limited by the state preparation fidelity of $80.2(0.8)\%$. As the atoms are efficiently connected to a photonic channel, our gate paves the way towards quantum networking with multiqubit nodes and the distribution of entanglement in repeater-based long-distance quantum networks.  
\end{abstract}
\maketitle

Quantum logic gates are basic building blocks of quantum information processing protocols \cite{nielsen2000,ladd2010}. They have been implemented with different qubit carriers like ions \cite{schmidt-kaler2003,leibfried2003}, atoms \cite{isenhower2010,saffman2010}, photons \cite{kok2010,hacker2016}, solid-state spins \cite{vandersar2012} and superconductors \cite{yamamoto2003}. For the construction of a large-scale quantum network \cite{kimble2008}, however, interfaces are needed that combine the processing capabilities of these quantum gates with an efficient connection to optical channels. An ideal platform for this is provided by neutral atoms in an optical cavity. The enhanced light-matter interaction in this system allows for a strong coupling of the atomic matter qubits to flying photonic qubits. The implementation of a quantum gate inside such a multiqubit node, as demonstrated here, has been a long-standing goal as it adds the capacity for local quantum information processing to the network nodes. 

Our gate demonstrates a functional quantum device that has distinct advantages. First, it employs a flying photon and is thus readily integrated in a distributed quantum network. Second, it requires only one physical step and is thus fast, $2\unit{\micro s}$ in our case. Third, the reflected photon acts as a herald to recoup experimental losses in the otherwise deterministic protocol \cite{duan2005,lin2006}. Last but not least, it is independent of the type of qubits and could also be implemented with ions, diamond-defect centers, superconducting qubits, or possibly even a combination of these. Our gate might therefore become a valuable tool in a future quantum networking architecture based on any of these platforms. For instance, it could serve as a Bell-state analyzer for swapping and distributing entanglement in a quantum repeater \cite{briegel1998} based on memory qubits in optical cavities \cite{uphoff2016}.

\begin{figure}
\centering
\hypertarget{fig:setup}{}
\includegraphics[width=7.4cm]{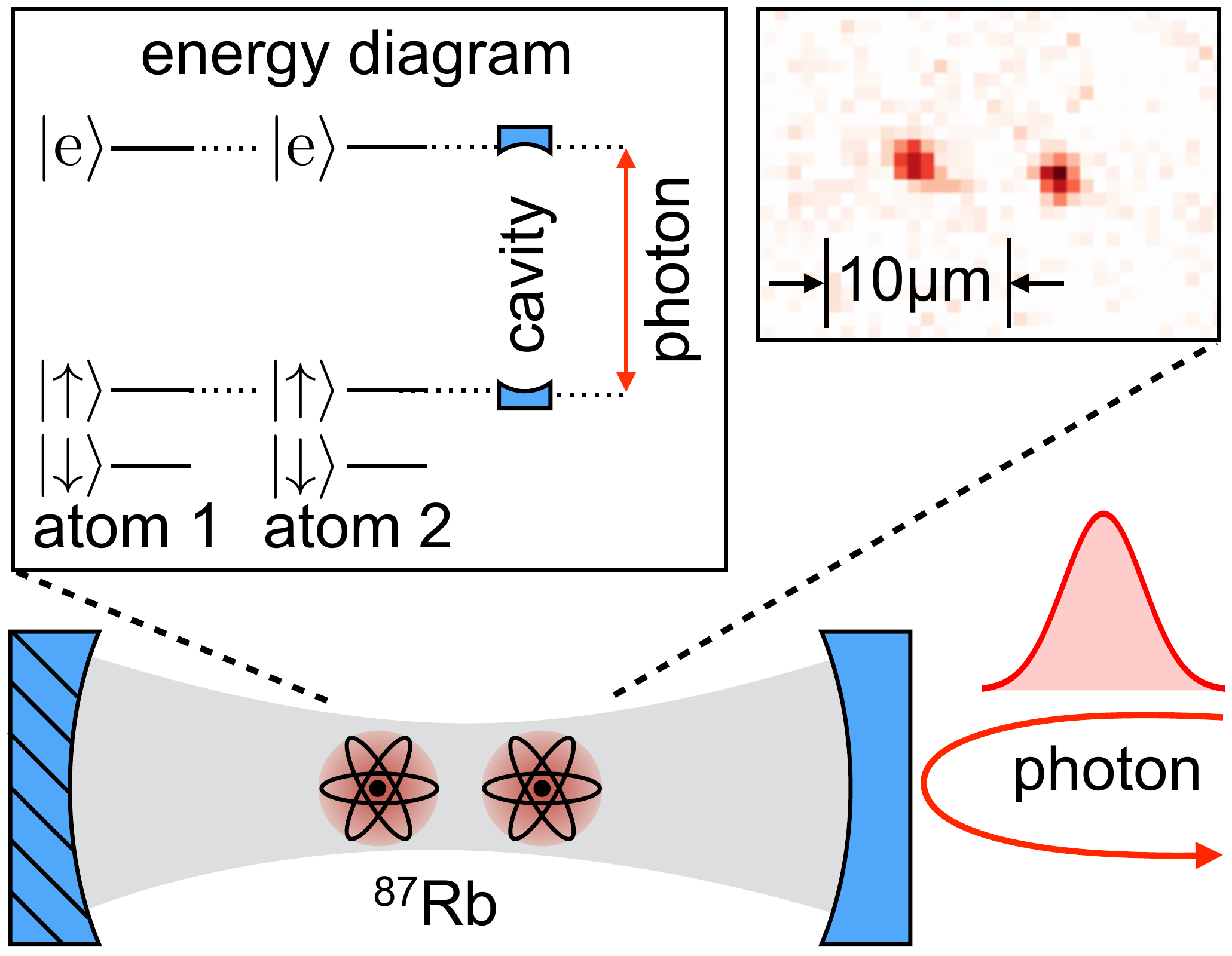}
\caption{\label{fig:setup}
Schematic representation of the setup. Two \textsuperscript{87}Rb atoms are trapped at the center of a cavity with asymmetric mirror transmissions. Photons impinge on the mirror with higher transmission and are reflected. The inset on the upper left-hand side shows a simplified level scheme of the atoms with the two ground states $\ket{\down}$, $\ket{\up}$ encoding the qubit and the excited state $\ket{\text{e}}$. The cavity is tuned to the $\ket{\up}\,{\leftrightarrow}\,\ket{\text{e}}$ transition. The inset on the upper right shows a fluorescence image of two trapped atoms in the cavity.}
\end{figure}

Our experiment follows proposals formulated more than a decade ago \cite{duan2005,lin2006}. The experimental setting, sketched in Fig.~\figref{fig:setup}, is an asymmetric high-finesse ($F=6{\times}10^4$) optical Fabry-P{\'e}rot cavity with one high-reflection mirror ($R=99.9994\%$) and one outcoupling mirror ($R=99.99\%$), through which flying photons can enter and leave.

Two \textsuperscript{87}Rb atoms are trapped at antinodes of the resonant cavity field and are individually observed with a microscope. The relevant cavity QED parameters are $(g,\kappa,\gamma)=2\pi\,(7.8,2.5,3)\unit{MHz}$, where $g$ denotes the single atom-cavity coupling rate, $\kappa$ the cavity field decay rate and $\gamma$ the atomic polarization decay rate \cite{reiserer2013}. These parameters place our atom-cavity system in the strong coupling regime \cite{reiserer2015}. Each of the atoms carries a qubit encoded in two stable hyperfine states $\ket{\up}=5^2S_{1/2}\ket{F{=}2,m_F{=}2}$ and $\ket{\down}=5^2S_{1/2}\ket{F{=}1,m_F{=}1}$. The atomic state $\ket{\up}$ is strongly coupled to the cavity light field via an optical transition $\ket{\up}\,{\leftrightarrow}\,\ket{\textnormal{e}}=5^2P_{3/2}\ket{F'{=}3,m_F{=}3}$, whereas $\ket{\down}$ is not. If none of the atoms couples to the light field, a resonant single photon reflected from the atom-cavity system can enter the resonator before leaving through the outcoupling mirror. This induces a $\pi$-phase shift in the combined atom-photon state \cite{lin2006,duan2005,duan2004,reiserer2013b,tiecke2014}. Any strongly coupled atom causes a normal-mode splitting, governed by the Tavis-Cummings Hamiltonian, such that an impinging photon does not fit spectrally into the cavity. In this case, the photon is directly reflected from the outcoupling mirror and the atom-photon state acquires no phase shift. This effect realizes a two-atom controlled-Z gate \cite{nielsen2000}:%

\begin{equation}
\label{eq:gate}
\begin{array}{rclcrcl}
\ket{\up  \up  }&\rightarrow&   \ket{\up  \up  }&\qquad&
\ket{\up  \down}&\rightarrow&   \ket{\up  \down}\\
\ket{\down\up  }&\rightarrow&   \ket{\down  \up}&\qquad&
\ket{\down\down}&\rightarrow& - \ket{\down\down}
%&\ .
\end{array}%
\end{equation}%

Qubit rotations are performed via a pair of Raman lasers which copropagate perpendicular to the cavity axis and illuminate both atoms equally. The Raman lasers are linearly polarized along and perpendicular, respectively, to the cavity axis which serves as a quantization axis. We perform arbitrary global qubit rotations by controlling the pulse duration, detuning and phase. The experimental sequences described hereafter are started after a successful loading of two atoms with distance $2\le d\le 12\unit{{\micro}m}$. This allows us to limit the spread of Rabi frequencies for different interatomic distances and thus increases the average fidelity of the gate. The center of mass of the two atoms is actively positioned to overlap with the center of the Raman beams. More details of the experimental setup are given in Appendix~\ref{sec:experimentalsetup}.

Experimentally, we quantify the performance of our two-atom quantum gate with two key operations which demonstrate its functionality: the creation of a maximally entangled output state from a separable input state as well as the controlled-NOT (CNOT) operation on a set of four basis states, in our case the Bell states.

At the start of each experiment, the atomic qubits are initialized through optical pumping and optionally subsequent coherent state rotations and quantum state carving \cite{soerensen2003a,welte2017}. In this way, we initially prepare $\ket{\down\down}$ or one of the four Bell states for our gate characterization measurements. Details of the inherently probabilistic carving technique for the preparation of the Bell states are given in Ref.~\cite{welte2017}. To execute the gate, we impinge a weak coherent pulse (mean photon number $\overline{n}=0.13$) and herald the presence of a photon through its detection with a conventional single-photon counter after the reflection.

Following each gate operation, the resulting two-atom state is read out by resonantly probing the cavity transmission and the atomic fluorescence, with a global atomic $\pi$ rotation in between. This allows us to distinguish between the states $\ket{\down\down}$, $\ket{\up\up}$ and the set $\{\ket{\up\down},\ket{\down\up}\}$ in each experiment. A detailed description of the state detection protocol can be found in Appendix~\ref{sec:entiregatesequence}.

Repeated measurements of identically prepared states yield the state populations $P_{ij}$ as probabilities to find the first atom in the state $\ket{i}$ and the second in $\ket{j}$ with $i,j\in\{\up,\down\}$. They constitute the diagonal elements of the two-atom density matrix $\rho$ as $\rho_{\up\up,\up\up}=P_{\up\up}$, $\rho_{\down\down,\down\down}=P_{\down\down}$ and $\rho_{\up\down,\up\down}{+}\rho_{\down\up,\down\up}=P_{\up\down}{+}P_{\down\up}$. Furthermore, the relevant off-diagonal elements of $\rho$ are obtained using the method of parity oscillations \cite{turchette1998,sackett2000}: To this end, the populations $\tilde P_{ij}(\phi)$ are determined after an additional $\pi/2$ pulse of varying phase $\phi$ relative to all previous rotation pulses. The parity $\Pi$, defined as $\Pi(\phi):=\tilde P_{\up\up}(\phi){+}\tilde P_{\down\down}(\phi)-[\tilde P_{\up\down}(\phi){+}\tilde P_{\down\up}(\phi)]$, is evaluated for each value of $\phi$. Analytically, $\Pi(\phi)=2\mathop{\mathrm{Re}}(\rho_{\up\down,\down\up})+2\mathop{\mathrm{Im}}(\rho_{\up\up,\down\down})\sin(2\phi)+2\mathop{\mathrm{Re}}(\rho_{\up\up,\down\down})\cos(2\phi)$ and, therefore, coherence terms of the density matrix can be extracted from a fit to the measured parity data. This set of parameters gives sufficiently many, linearly independent parameters to determine the fidelities of the produced states with the four states of a Bell basis \cite{sackett2000,ospelkaus2011} 
\begin{equation}
\label{eq:bellfideliy}
\begin{array}{rcl}
F(\ket{\Psi^\pm})&=&\textstyle\frac12(P_{\up\down}+P_{\down\up})\pm\mathop{\mathrm{Re}}(\rho_{\up\down,\down\up}),\\[1.5ex]
F(\ket{\Phi^\pm})&=&\textstyle\frac12(P_{\up\up}+P_{\down\down})\pm\mathop{\mathrm{Re}}(\rho_{\up\up,\down\down}).
\end{array}
\end{equation}

\begin{figure}
\centering
\hypertarget{fig:parity}{}
\includegraphics[width=8cm]{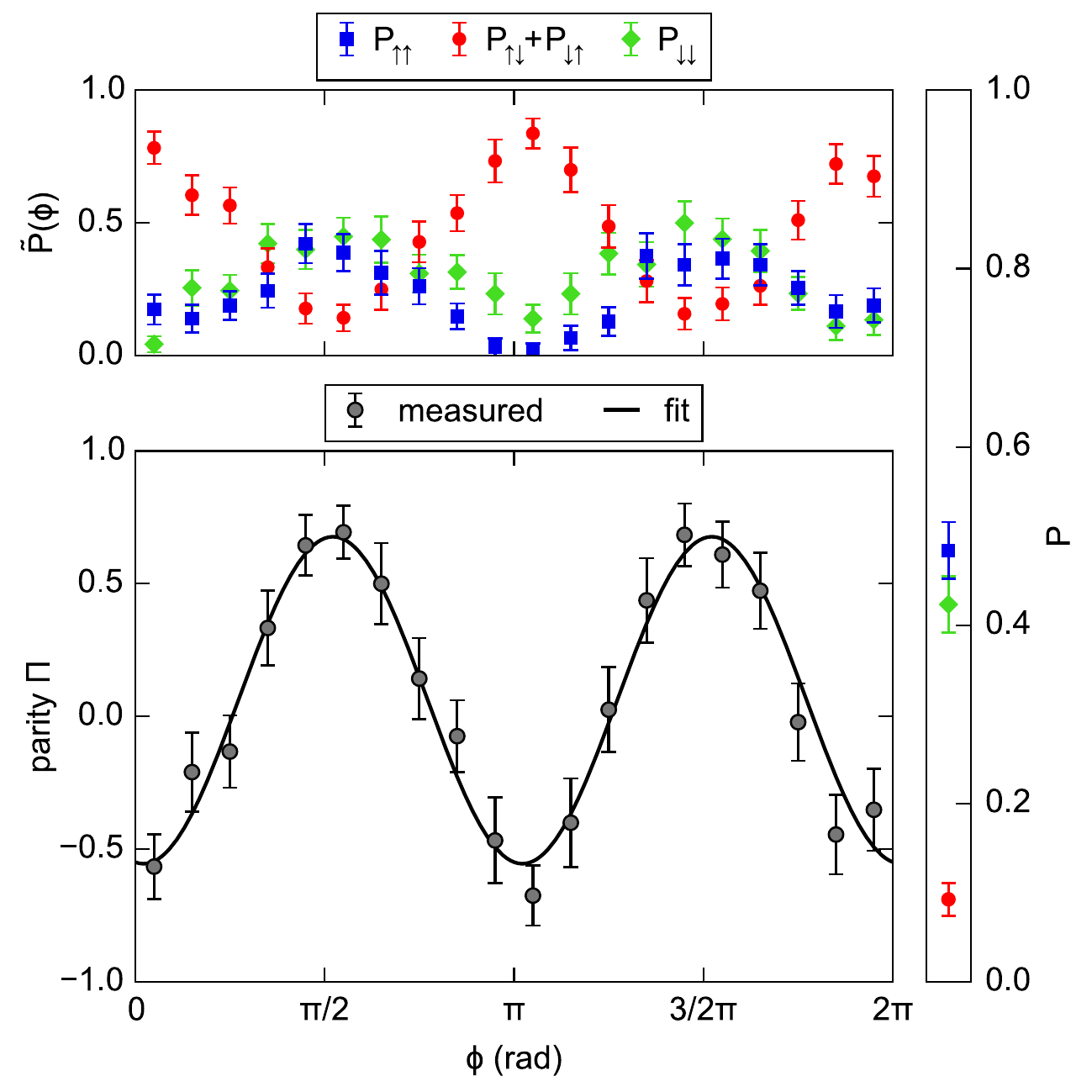}
\caption{\label{fig:parity}
Parity oscillations and populations of the entangled state produced by the gate and a subsequent $\pi/4$ rotation. The populations $\tilde P_{\up\up}$ (blue squares), $\tilde P_{\up\down}+\tilde P_{\down\up}$ (red circles) and $\tilde P_{\down\down}$ (green diamonds) oscillate as a function of the phase of the analysis pulse $\phi$, which is scanned from $0$ to $2\pi$. They yield the parity $\Pi(\phi)$, which we fitted with a sine of variable offset, phase and amplitude. The values for the populations $P$ on the right-hand side come from a measurement without an analysis pulse. The overlap with the maximally entangled state $\ket{\Phi^-}$ inferred from this data is $F=76(2)\%$. Error bars are statistical standard deviations of the mean within each bin. The slight offset in the phase is due to a residual two-photon Raman-laser detuning of ${\pm}3\unit{kHz}$.}
\end{figure}

We demonstrate an entangling gate operation on the separable input state $\frac12\left(\ket{\up}-\ket{\down}\right)\otimes\left(\ket{\up}-\ket{\down}\right)=\frac12\left(\ket{\up\up}-\ket{\up\down}-\ket{\down\up}+\ket{\down\down}\right)$, which we prepare from $\ket{\down\down}$ by a collective $\pi/2$ rotation. The gate flips the sign of the $\ket{\down\down}$ component (Eq.~(\ref{eq:gate})) and thus immediately creates a maximally entangled state. This becomes apparent by applying a further global state rotation with a pulse area of $\pi/4$, resulting in the Bell state $\ket{\Phi^-}=\frac{1}{\sqrt{2}}\left(\ket{\up\up}-\ket{\down\down}\right)$. In the experiment, we obtain a fidelity of $F(\ket{\Phi^-})=76(2)\%$. The state populations and the parity are shown in Fig.~\figref{fig:parity}, with populations of the produced output state of $P_{\up\up}=48(3)\%$, $P_{\up\down}+P_{\down\up}=9(2)\%$ and $P_{\down\down}=42(3)\%$ and a fit to the parity oscillation data giving $\mathop{\mathrm{Re}}(\rho_{\up\up,\down\down})=-31(2)\%$.

The entangling operation shows that the gate has an entangling capability \cite{poyatos1997} of $\mathcal C\leq-0.26(2)$ (statistical), given by the smallest absolute negativity of any possible state with our determined density matrix entries. An ideal entangling gate has $\mathcal{C}=-0.5$, whereas any nonentangling gate has $\mathcal{C}=0$.

\begin{table}[b]
\caption{\label{tab:errors}%
Sources of error for the experimental fidelity of the $\ket{\Phi^-}$ state created by the entangling gate operation. Given values are the absolute decrease in fidelity due to each effect if all other influences are kept constant. The effects are not independent, which has to be taken into account if multiple error sources are summed.
}
\begin{ruledtabular}%
\begin{tabular}{lc}%
Source of error & Fidelity reduction \\
\hline
Finite mode matching & 6\% \\
Erroneous state detection & 4\% \\
Multiphoton contributions & 3\% \\
Photon loss in the cavity & 3\% \\
Heralding detector dark counts & 2\% \\
Photon polarization inaccuracy & 1\% \\
Atomic state preparation & 1\% \\
Atomic state dephasing & 1\% \\
\end{tabular}%
\end{ruledtabular}%
\end{table}
The fidelity's deviation from unity is well explained by known error sources. To this end, we model the atom-cavity system using input-output theory and treated the two-atom state using its density matrix (see Appendix~\ref{sec:theory}). The calculation shows the effect of the different experimental imperfections on the measured fidelity. They are listed in Table~\ref{tab:errors}. The biggest error of 6\% stems from an imperfect transversal mode matching (92\%) of the impinging photon to the cavity mode. In the noncoupling case, the phase flip in the atom-atom-photon state does not occur and no entanglement is generated despite the heralding event. The second-most important source of error is the final state detection. The Poissonian distributions of detected photons for coupled and noncoupled atoms, respectively, have an overlap of 2\% for the state detection in transmission and 3\% for the state detection in fluorescence. With the two required state-detection intervals this error enters twice and causes a 4\% reduction in fidelity. The use of coherent pulses instead of single photons creates an 8\% chance to have two or more photons reflected despite detecting only one. Then the second photon reverses the gate and reduces the total fidelity by 3\%. The cavity itself has only a finite cooperativity, which results in a finite probability of photon scattering from the mirrors or atoms or transmission through the cavity instead of backreflection, thereby leaking information about the atomic state into the environment. This is the most fundamental source of errors, which is given by our choice of atomic states and mirrors. It causes a reduction in fidelity by 3\%. Other sources of error such as dark counts, dephasing, state preparation, fluctuating parameters, and inaccuracies are in the 1\% range.

The second benchmark measurement is given by the operation as a CNOT logic gate. Such a gate leaves two of the four input basis states unaffected, whereas the other two are interchanged. A conventional choice to perform this measurement is the basis $\{\ket{\up{\rightarrow}},\allowbreak \ket{\up{\leftarrow}},\allowbreak \ket{\down{\rightarrow}},\allowbreak \ket{\down{\leftarrow}}\}$, where $\ket{\rightarrow}=\frac{1}{\sqrt{2}}(\ket{\up}+\ket{\down})$ and $\ket{\leftarrow}=\frac{1}{\sqrt{2}}(\ket{\up}-\ket{\down})$. However, we want to employ a simple addressing scheme that avoids the necessity to discriminate between the two atoms during state preparation and state detection. Therefore, we use the global Bell basis states {$\ket{\Psi^-}, \ket{\Psi^+}, \ket{\Phi^-}, \ket{\Phi^+}$} as an input. In this basis, the gate operation swaps the states $\ket{\Phi^-}$ and $\ket{\Phi^+}$:
\begin{equation}
\label{eq:truthtable}
\begin{array}{rcl}
\ket{\Psi^{-}}=\textstyle\frac{1}{\sqrt{2}}\big(\ket{\up\down}-\ket{\down\up}\big)
&\rightarrow&
\textstyle\frac{1}{\sqrt{2}}\big(\ket{\up\down}-\ket{\down\up}\big)=\ket{\Psi^{-}}\\[1.5ex]
\ket{\Psi^{+}}=\textstyle\frac{1}{\sqrt{2}}\big(\ket{\up\down}+\ket{\down\up}\big)
&\rightarrow&
\textstyle\frac{1}{\sqrt{2}}\big(\ket{\up\down}+\ket{\down\up}\big)=\ket{\Psi^{+}}\\[1.5ex]
\ket{\Phi^{-}}=\textstyle\frac{1}{\sqrt{2}}\big(\ket{\up\up}-\ket{\down\down}\big)
&\rightarrow&
\textstyle\frac{1}{\sqrt{2}}\big(\ket{\up\up}+\ket{\down\down}\big)=\ket{\Phi^{+}}\\[1.5ex]
\ket{\Phi^{+}}=\textstyle\frac{1}{\sqrt{2}}\big(\ket{\up\up}+\ket{\down\down}\big)
&\rightarrow&
\textstyle\frac{1}{\sqrt{2}}\big(\ket{\up\up}-\ket{\down\down}\big)=\ket{\Phi^{-}}
\end{array}%
\end{equation}%

\begin{figure}[tb]
\centering
\hypertarget{fig:truthtable}{}
\includegraphics[width=8cm]{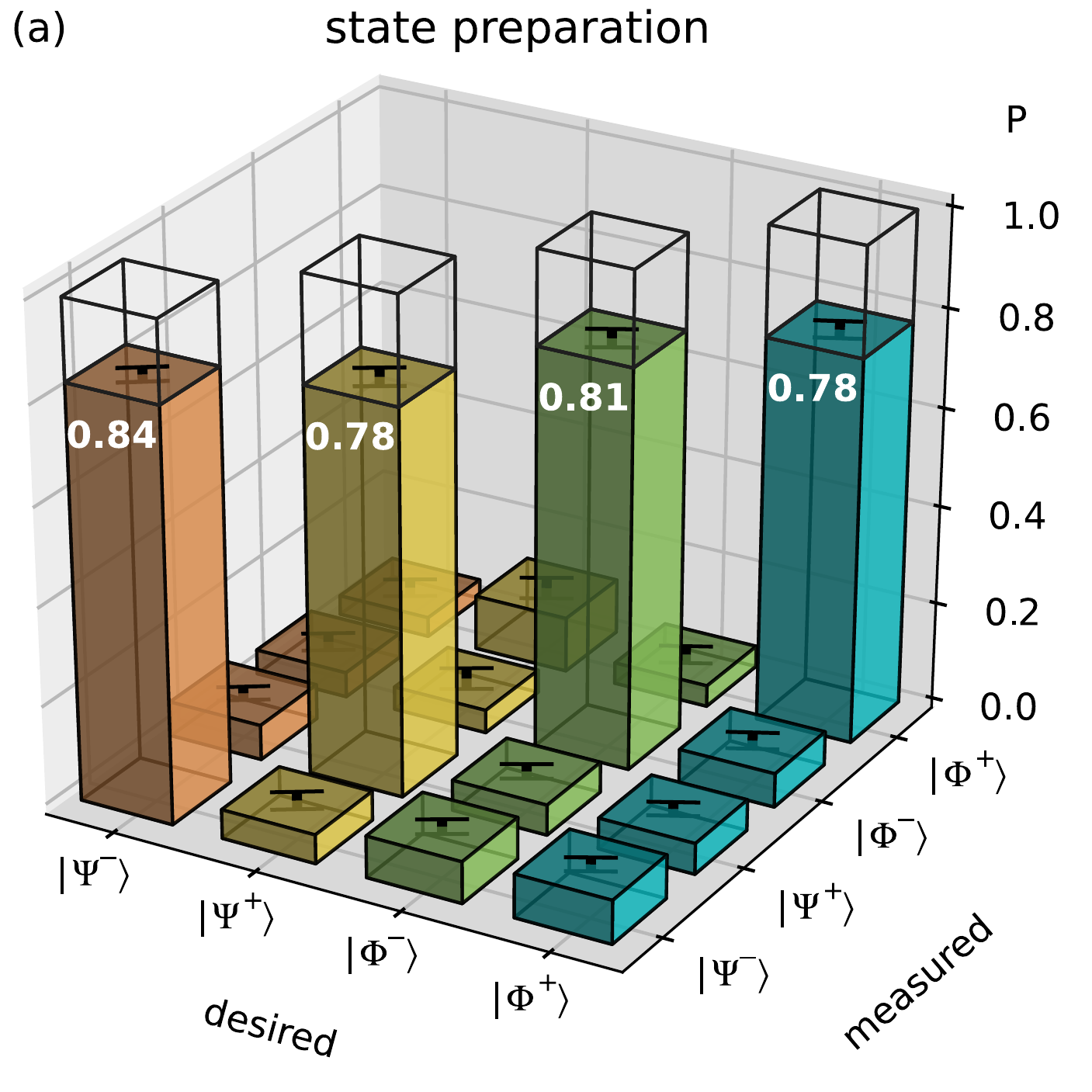}
\includegraphics[width=8cm]{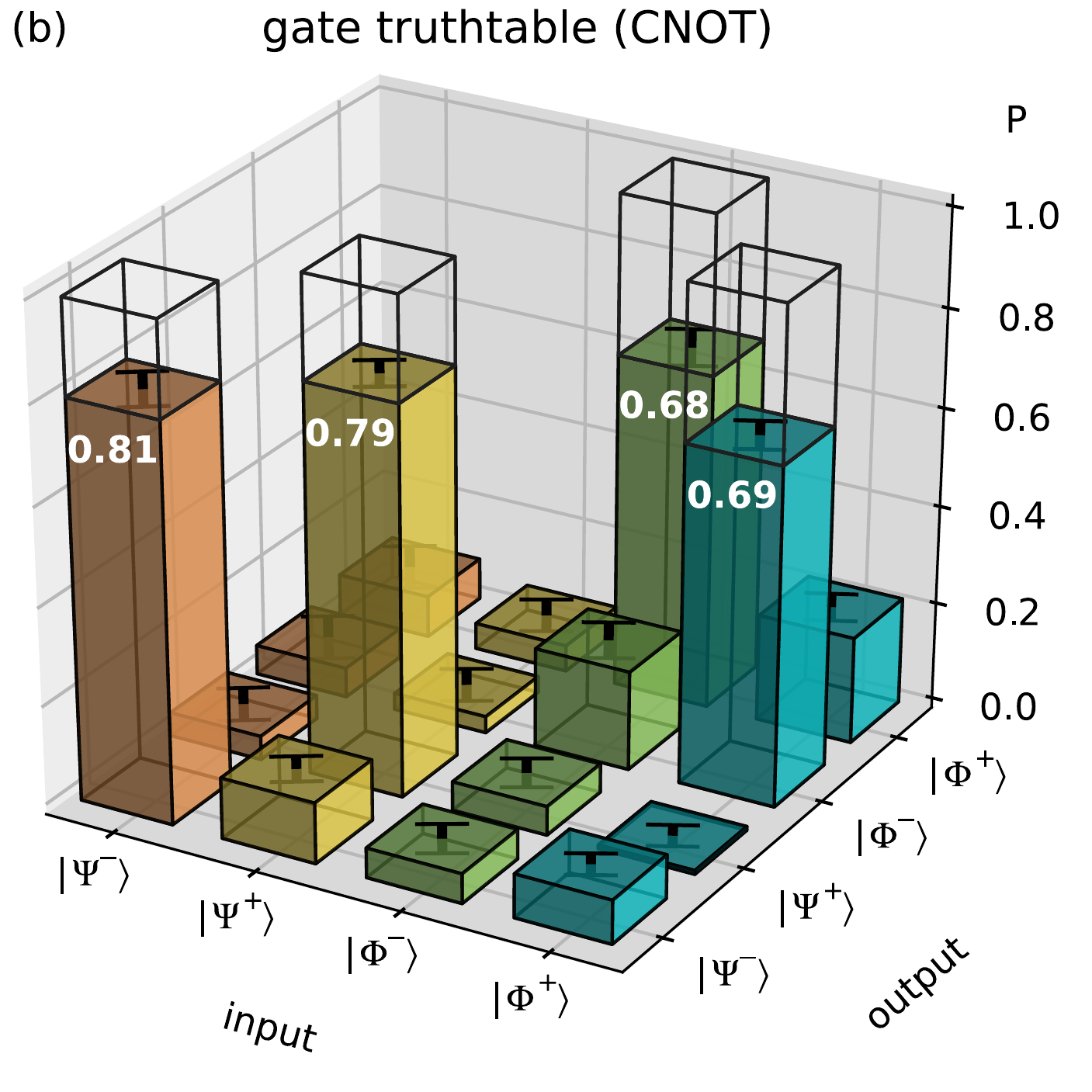}
\caption{\label{fig:truthtable}\label{fig:truthtableb}
Fidelities of the experimentally generated states in the Bell basis. (a) Bar diagram of the fidelities of the four Bell states used as an input for our gate. (b) Measured fidelities of the two-atom states resulting from the gate operation on these input states. Error bars are statistical standard deviations of the mean. The truth tables for ideal input states and gate operations are shown as transparent bars for comparison.}
\end{figure}

Using the Bell states does not only prove the ability of our gate to act as a CNOT but also demonstrates its capability to accept entangled input states and to preserve their quantum features. Experimentally, we prepare each of the four input Bell states with the method of quantum state carving \cite{soerensen2003a,welte2017} and achieve an average state preparation fidelity of $80.2(0.8)\%$ [Fig.~\figref{fig:truthtable}(a)]. To perform the carving operation, we initially pump the atoms into a parallel ($\ket{\down\down}$) or an antiparallel configuration ($\ket{\down\up}, \ket{\up\down}$). The initial pumping into the parallel configuration allows us to generate the triplet states $\ket{\Phi^+}$, $\ket{\Phi^-}$, or $\ket{\Psi^+}$ while the antiparallel initial states allow us to generate the singlet state $\ket{\Psi^-}$. After the pumping stage, both atoms are transferred into an equal superposition state with a $\pi/2$ pulse. We then reflect an antidiagonally polarized photon from the cavity, which changes its polarization to diagonal if there is at least one coupling atom in the cavity. A polarization resolved measurement of this photon after the reflection allows us to remove (i.e., carve) the $\ket{\down\down}$ component from the combined atom-atom state. Two of these carving steps enable the generation of all maximally entangled states to be used as an input for our quantum gate operation. Having performed the gate via the reflection of the gate-triggering photon, the output is again characterized in the Bell basis [Fig.~\figref{fig:truthtable}(b)] through parity oscillations (Appendix~\ref{sec:truthtablemethods}). The combined sequence for this measurement is described in more detail in Appendix~\ref{sec:entiregatesequence}.

From the experiment, we find an average overlap with an ideal CNOT truth table of $74.1(1.6)\%$, which is $92.3(2.2)\%$ of the input state fidelity. For $\ket{\Psi^+}$ and $\ket{\Psi^-}$, the output does not differ significantly from the input and the gate behaves as the expected identity operator. In contrast, the gate performs the anticipated NOT operation on $\ket{\Phi^+}$ and $\ket{\Phi^-}$, which is an exchange of their populations. Similarly to the entangling gate operation, the reduced fidelity after the CNOT operation mainly results from two-photon contributions in the impinging light pulse and from gate photons that did not match the cavity mode. 

In 4.2\% of all experimental runs, we detect one photon in the reflected weak coherent laser pulse ($\overline{n}=0.13$). The probability to lose a photon in the gate process is determined by the reflectivity of the cavity on resonance. In both the coupling and the noncoupling case, this reflectivity amounts to $67\%$ leading to a loss of $33\%$ of the photons. The reflected photons are detected with an efficiency of $55\%$. This detection is used to herald the gate operation. The probability to pump an atom into a state outside of the qubit state space can be approximated as $2.4\times 10^{-4}$ and mainly results from polarization imperfections due to cavity birefringence and off-resonant pumping of the atom via the $\ket{F'{=}3,m_F{=}1,2}$ states. Since loss of the trapped atoms is virtually absent (probability per gate operation $2\times 10^{-5}$) on the time scale of the full experimental sequence (${\approx}\,70\unit{{\micro}s}$), the overall gate efficiency amounts to the same $4.2\%$.

Unity efficiency could be achieved with an external single-photon source, thus avoiding the need to postselect on a heralding event. For such an ideal input, with all other parameters left the same, we infer from our simulation in Appendix~\ref{sec:theory} that the gate would still achieve a fidelity of $68\%$. Our cavity system is then able to perform deterministic gate operations and create entangled atom pairs on demand. The fidelity reduction by $8\%$ mainly stems from the possibility to lose the photon in the cavity and transport information about the individual atomic states into the environment. On the positive side, the loss of fidelity is mitigated by the absence of zero- and multiphoton events. Thus, if a single-photon source is used together with heralding, the fidelity would get as high as $82\%$ at a heralding efficiency of $32\%$, much higher than with the weak-coherent input.

To conclude, we demonstrate the quantum functionality of a novel two-atom gate that is mediated by a single optical photon. The gate can be readily extended to implement hybrid atom-atom-photon gates such as the Toffoli, since the photon can carry a polarization qubit with a state-dependent atom-photon interaction strength \cite{reiserer2015}. The long-range nature of the mediating optical field allows for an extension to one-step multiqubit gates on several atoms simply by placing more atoms into the same cavity mode \cite{lin2006}. This could lead to further applications like an error-correction scheme with three atoms \cite{borregaard2015}. With an additional single-atom addressing system to shift individual qubits in and out of resonance, our gate could function as a building block for larger quantum circuits within one cavity. Individual atomic qubits from a larger ensemble could further be mapped from and onto flying photons with the scheme of Ref.~\cite{kalb2015}, with one atom tuned into resonance during each reflection.

As the gate is implemented in a system that is ready to serve as a quantum network node \cite{ritter2012,northup2014}, it provides a way towards distributed quantum computing through photonic links.

Our demonstrated gate mechanism does not rely on the specific platform of neutral \textsuperscript{87}Rb, but could equally well be applied to many other types of natural and artificial atoms. For example, it could be implemented with superconducting qubits inside microwave cavities \cite{majer2007} where many of the required techniques are available. This could establish a new protocol for small-scale quantum networks in solid-state systems, with microwave photons released from one node executing gates in potentially several others.

\vspace{\baselineskip}
\section*{\normalfont\bfseries ACKNOWLEDGMENTS}
We thank S.\ D\"urr, P.\ Eder, M.\ Fischer, M.\ K\"orber, A.\ Neuzner, A.\ Reiserer and M.\ Uphoff for valuable ideas and discussions. This work was supported by the Bundesministerium f\"ur Bildung und Forschung via IKT 2020 (Q.com-Q) and by the Deutsche Forschungsgemeinschaft via the excellence cluster Nanosystems Initiative Munich (NIM). S.W. was supported by {\em Elitenetzwerk Bayern} (ENB) through the doctoral program ExQM.

%\clearpage
\appendix
\titleformat{\section}[block]{\normalfont\bfseries\centering}{APPENDIX \thesection:}{0.5em}{}
\titleformat{\subsection}[block]{\normalfont\bfseries\centering}{\thesubsection.}{0.5em}{}

\section{EXPERIMENTAL SETUP}\label{sec:experimentalsetup}
Two \textsuperscript{87}Rb atoms are trapped at the center of a single-sided, high-finesse ($F=6{\times}10^4$) cavity of length $0.5\unit{mm}$ and mode waist $30\unit{{\micro}m}$. The atoms are trapped in a three-dimensional optical lattice potential \cite{reiserer2013} and cooled to temperatures of ${\approx}\,100\unit{{\micro}K}$ via a Sisyphus cooling technique. Employing a piezo crystal, the cavity is tuned into resonance with the atomic D$_2$ transition $5^2S_{1/2}\ket{F{=}2,m_F{=}2}\leftrightarrow5^2P_{3/2}\ket{F'{=}3,m_F{=}3}$ at $780\unit{nm}$. We encode the atomic qubits in the hyperfine ground states $\ket{\up}=5^2S_{1/2}\ket{F{=}2,m_F{=}2}$ and $\ket{\down}=5^2S_{1/2}\ket{F{=}1,m_F{=}1}$. The two qubits are controlled via a pair of Raman lasers which drive the $\ket{\up}\leftrightarrow\ket{\down}$ transition coherently. Both beams impinge from the same direction, orthogonally to the cavity axis. Their beam waist of $w_0=35\unit{{\micro}m}$ is much bigger than the interatomic distance $d$, such that both atoms are controlled identically. 

We have two different projective readout techniques, which are used for state preparation and state detection. Both yield the same information, namely whether at least one atom couples to the cavity, or none ($\ket{\down\down}$). One technique is probing the cavity transmission through the high-reflectivity mirror. The transmission of light resonant with the $\ket{\up}\,{\leftrightarrow}\,\ket{\text{e}}$ transition is strongly suppressed if any atom is in the coupling state $\ket{\up}$ compared to $\ket{\down}$, due to a normal-mode splitting. In the second technique, we illuminate the atoms with a resonant beam impinging orthogonally to the cavity axis. Any atom in $\ket{\up}$ will scatter fluorescence photons into the cavity mode and the light can be collected with single-photon counters. The fluorescence technique yields a readout with higher fidelity, but it can remove atoms in $\ket{\up}$ from the qubit subspace, which prohibits its application at intermediate stages of the sequence.

\section{SEQUENCE OF THE EXPERIMENT}\label{sec:entiregatesequence}
Figure~\figref{fig:entiregate} shows a quantum circuit diagram of the full experimental sequence. In order to characterize the gate, which consists of only one experimental step, we need to provide several defined input states and measure the gate output in different bases. Here, we explain the various steps which are part of the experiment, additionally to the gate itself.\vspace{0.5\baselineskip}

\subsection{Atomic state preparation}
\textsuperscript{87}Rb has several stable 5\textsuperscript{2}S$_{1/2}$ ground states, which are outside of the qubit space but may be populated at the beginning of an experiment. Therefore, we start with optical pumping of the two atoms into the qubit manifold using resonant, right-circularly polarized laser light along the cavity axis, which also is our quantization axis defined by a small magnetic guiding field. The state $\ket{\up}=\ket{F{=}2,m_F{=}2}$ connects only to a cycling transition and thus accumulates the whole population. However, once an atom is in $\ket{\up}$, it strongly reduces the intracavity pump light via a normal-mode splitting and will hamper the preparation of a second atom. To realize a state with both atoms in a parallel state $\ket{\up\up}$ nevertheless, we detune the atomic resonance from the cavity, using the dynamical Stark effect of the trapping laser during the optical pumping process. To herald a successful preparation of both atoms, we employ a global $\pi$ pulse, perform a state detection in fluorescence and postselect on those cases in which no fluorescence photons are detected. This results in an effective preparation of the state $\ket{\down\down}$. The incoherent mixture of two-atom states with density matrix $\rho=\frac{1}{2}(\ket{\up\down}\bra{\up\down}+\ket{\down\up}\bra{\down\up})$ can be prepared by resonantly pumping a first atom to the strongly coupled state $\ket{\up}$, followed by a $\pi$ pulse to $\ket{\down}$ and another resonant pumping sequence for the second atom. The proper preparation of this incoherent mixture is verified by two state detections in transmission with an interleaved $\pi$ pulse. If a low transmission is observed in both state-detection intervals, the antiparallel state preparation has been successful.

\subsection{Bell basis generation}
The next step in the experiment is the preparation of Bell states, which we use to demonstrate the CNOT operation. Starting from an initial preparation of the state $\ket{\down\down}$, the triplet Bell states $\ket{\Psi^+}$ and $\ket{\Phi^\pm}$ can be generated by quantum state carving \cite{welte2017}. For this, two antidiagonally polarized photons (state $\ket{\text{A}}$) are subsequently reflected from the cavity with a global $\pi$ rotation of the atoms in between. A photon polarization change to $\ket{\text{D}}$ (diagonal) heralds that at least one of the atoms is in $\ket{\up}$ and therefore couples to the cavity. A postselection on the flipped cases carves $\ket{\Psi^+}$ out of an initially separable coherent spin state. Afterwards, global $\pi/2$ pulses around different axes are used to generate $\ket{\Phi^\pm}$ states from $\ket{\Psi^+}$. When starting with an antiparallel mixture initially, the same sequence generates the $\ket{\Psi^-}$ singlet state. In contrast to the quasideterministic entangling operation of our demonstrated gate, this preparation of Bell basis states through carving is inherently probabilistic.

After the gate input $\ket{\Psi^{\text{In}}}$ preparation, the atom-atom gate operation is executed by reflecting a right-circularly polarized photon from the cavity, as explained in the main text. The gate entangles a separable $\frac{1}{2}(\ket{\up\up}-\ket{\up\down}-\ket{\down\up}+\ket{\down\down})$ input, leaves the $\ket{\Psi^\pm}$ states untouched and interchanges the $\ket{\Phi^\pm}$ states.

\subsection{State analysis rotation}
The gate output $\ket{\Psi^{\text{Out}}}$ needs to be analyzed for populations and coherences. To access the latter, we employ the method of parity oscillations \cite{turchette1998,sackett2000}, as explained in the main text, where an optional analysis pulse of pulse area $\pi/2$ with a variable phase of $\phi$ is applied on both qubits identically. This rotation turns coherence terms into populations $\tilde P_{ij}(\phi)$, which can be evaluated by fitting of a theoretical model.

\subsection{State detection}
The final readout of the two-atom output state is performed by a sequence of two state detection pulses with an interleaved $\pi$-rotation. The first state detection probes the cavity transmission and allows to distinguish $\ket{\down\down}$ from $\{\ket{\up\down},\ket{\down\up},\ket{\up\up}\}$, as states with at least one coupling atom lead to a strong reduction of the cavity transmission. Employing a $\pi$-rotation, an initial $\ket{\up\up}$ state is subsequently transferred into $\ket{\down\down}$ and can be discriminated from the set of remaining states via the second state detection performed in fluorescence. This double state detection protocol allows to distinguish the states $\ket{\down\down}$, $\ket{\up\up}$ and the set $\{\ket{\up\down},\ket{\down\up}\}$.

\onecolumngrid
\par\par % two newlines needed here
\begin{widetext}
\begin{figure}[h]
\centering
\hypertarget{fig:entiregate}{}
\includegraphics[width=150mm]{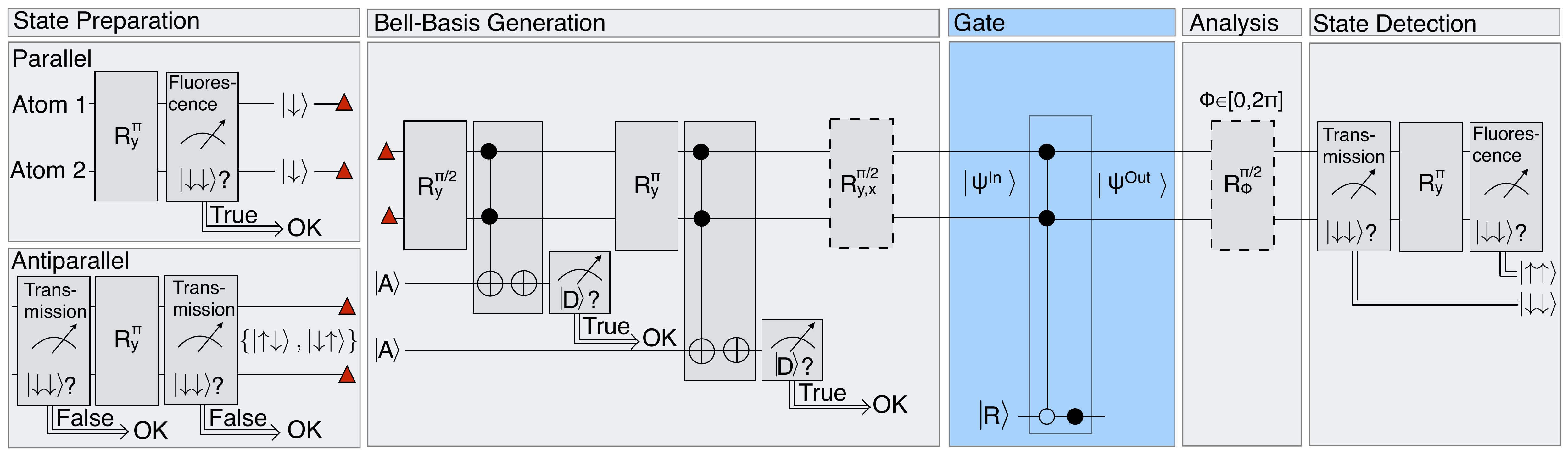}
\caption{\label{fig:entiregate}%
Complete quantum circuit diagram depicting initialization, preparation of Bell basis states, the two-atom gate, analysis state rotation and final state detection. Rotations (R) have a subscript indicating the rotation axis $x$ ($\ket{\up}+\ket{\down}$) or $y$ ($\ket{\up}+i\ket{\down}$) and a superscript denoting the rotation angle. The two techniques for state detection are labelled transmission and fluorescence, respectively. Whenever the result of an atomic state detection or a photon polarization measurement does not yield the desired result (OK), the protocol is restarted from the beginning.  Initialization with parallel states allows to prepare the triplet Bell states $\ket{\Psi^+}$ and $\ket{\Phi^\pm}$ as gate input. The singlet state $\ket{\Psi^-}$ is prepared after initialization with an incoherent mixture of antiparallel states described by a density matrix $\frac{1}{2}(\ket{\up\down}\bra{\up\down}+\ket{\down\up}\bra{\down\up})$. The photon reflection with an antidiagonal ($\ket{\text{A}}$) polarization realizes a Toffoli gate with an inversion of the photonic qubit. Experimentally, this means that the polarization of the reflected photon is flipped to $\ket{\text{D}}$ in all cases except the one where both atoms occupy $\ket{\down}$. The dashed box in the Bell basis generation section is optional and only needed when transforming $\ket{\Psi^+}$ to $\ket{\Phi^\pm}$. Various two-atom states $\ket{\Psi^{\text{In}}}$ are prepared as gate input, then processed by the two-atom gate and the result $\ket{\Psi^{\text{Out}}}$ is characterized by state detection in a two-atom basis set by a preceding analysis rotation. The dashed box in the analysis section is optional. For a  measurement of the populations $P_{i,j}$ it is omitted while for a measurement of the populations $\tilde P_{i,j}$, with $i,j\in\{\down,\up\}$, it is applied.}
\end{figure}
\end{widetext}
\twocolumngrid

\subsection{Timing}
A $\pi$ pulse in the experiment has a duration of $8\unit{{\micro}s}$. The photons have a Gaussian intensity profile with a full width at half maximum of $0.9\unit{{\micro}s}$, such that the phase-shift mechanism works reliably (see supplementary material of ref.~\cite{kalb2015} for details). We work with an average impinging photon number of $\overline{n}=0.33$ for each of the two $\ket{\text A}$ pulses in the state preparation, which was found to lead to high fidelities in the state-preparation \cite{welte2017}. For the right-circularly polarized gate photon $\ket{\text R}$, we use a lower average photon number of $\overline{n}=0.13$ to further suppress multiphoton contributions. The state-detection intervals have a length of $5\unit{{\micro}s}$ and $3\unit{{\micro}s}$ for a state detection in fluorescence and a state detection in transmission, respectively (see Fig.~\figref{fig:entiregate}). We repeat the protocol of length ${\approx}\,70\unit{{\micro}s}$ at a rate of $1\unit{kHz}$ with laser cooling intervals of $0.7\unit{ms}$. The actual time to perform the gate is given by the length of the gate photon whose full width at half maximum is below one microsecond.

\section{THEORETICAL CALCULATION OF THE GATE FIDELITY}\label{sec:theory}
The gate operation can be modeled using the density matrix formalism. First we describe the effect of one reflected photon on the atomic density matrix and then we add all other imperfections of the experiment.

\subsection{Effect of the photon reflection}
A photon $\ket{p_{\textnormal{in}}}$ that impinges in a cavity mode with $N$ coupling atoms, can end up in several orthogonal modes, namely reflection along the original mode $\ket{p_r}$, transmission through the cavity $\ket{p_t}$, scattering at the mirrors $\ket{p_m}$ and scattering at an atom number $i$ $\ket{p_{ai}}$. The resulting wave-function amplitudes $r$, $t$, $m$ and $a$ for each case are described by cavity input-output theory \cite{kuhn2015} using our measured cavity parameters $(g,\kappa,\gamma)=2\pi(7.8, 2.5, 3.0)\unit{MHz}$, and the field decay rates $\kappa_r=2\pi\cdot2.29\unit{MHz}$ through the incoupling mirror, $\kappa_t=2\pi\cdot0.09\unit{MHz}$ through the cavity back mirror and $\kappa_m=2\pi\cdot0.13\unit{MHz}$ from mirror scattering with $\kappa_r+\kappa_t+\kappa_m=\kappa$:
\begin{eqnarray}
r(N) &=& \frac{Ng^2+(\kappa-2\kappa_r)\gamma}{Ng^2+\kappa\gamma}
\quad(=-0.82, 0.80, 0.89)
\\
t(N) &=& \frac{2\sqrt{\kappa_r\kappa_t}\gamma}{Ng^2+\kappa\gamma}
\quad(=0.36, 0.04, 0.02)
\\
m(N) &=& \frac{2\sqrt{\kappa_r\kappa_m}\gamma}{Ng^2+\kappa\gamma}
\quad(=0.43, 0.05, 0.03)
\\
a(N) &=& \frac{2\sqrt{\kappa_r\gamma}\sqrt{N}g}{Ng^2+\kappa\gamma}
\quad(=0.00, 0.60, 0.45)
\end{eqnarray}
These amplitudes fulfill the normalization requirement $r(N)^2+t(N)^2+m(N)^2+a(N)^2=1$. Values in parentheses are given for $N=0,1,2$ coupled atoms, respectively.

We consider the atoms in the canonical basis of coupling ($\up$) and non-coupling ($\down$) states $\ket{a_0}=\ket{\up\up}$, $\ket{a_1}=\ket{\up\down}$, $\ket{a_2}=\ket{\down\up}$ and $\ket{a_3}=\ket{\down\down}$. When the atoms are prepared in an initial two-atom state $\rho_a$, the reflection of one photon in $\ket{p_{\textnormal{in}}}$
\begin{equation}
\rho_a\otimes\ket{p_{\textnormal{in}}}\bra{p_{\textnormal{in}}} = \sum_{i,j=0}^3 a_{i,j}\ket{a_i}\bra{a_j}\otimes\ket{p_{\textnormal{in}}}\bra{p_{\textnormal{in}}}
\end{equation}
results in a final state
\begin{equation}
\rightarrow \rho_{a,p} = \sum_{i,j=0}^3 a_{i,j}\ket{a_i}\bra{a_j}\otimes\ket{p_{i}}\bra{p_{j}}
\end{equation}
with
\begin{eqnarray}
\ket{p_{0}} &=& r(2)\ket{p_r} + t(2)\ket{p_t} + m(2)\ket{p_m} +\frac{a(2)}{\sqrt{2}}(\ket{p_{a1}} + \ket{p_{a2}})
\nonumber\\
\ket{p_{1}} &=& r(1)\ket{p_r} + t(1)\ket{p_t} + m(1)\ket{p_m} + a(1)\ket{p_{a1}}
\nonumber\\
\ket{p_{2}} &=& r(1)\ket{p_r} + t(1)\ket{p_t} + m(1)\ket{p_m} + a(1)\ket{p_{a2}}
\nonumber\\
\ket{p_{3}} &=& r(0)\ket{p_r} + t(0)\ket{p_t} + m(0)\ket{p_m}\ .
\end{eqnarray}
Generally, this constitutes an atom-atom-photon entangled state, but the photon gets absorbed immediately afterwards. In the case that a reflected photon is detected in $\ket{p_r}$ (heralding), the final atomic state is given by
\begin{eqnarray}
\rho_{a,\textnormal{out}}^{\textnormal{herald}}\cdot\mathcal{N} &=& \bra{p_r}\rho_{a,p}\ket{p_r}
\nonumber\\
&=& \rho_a\circ(\pmb{(}r(2), r(1), r(1), r(0)\pmb{)}^\dagger\cdot
\nonumber\\
&\ & \qquad\ \,\pmb{(}r(2), r(1), r(1), r(0)\pmb{)})\nonumber\\
&=:&\rho_a\circ G_{\textnormal{herald}}\nonumber\\
\label{eq:operator_herald}
&=& \rho_a\circ
\begin{pmatrix}
0.80 & 0.71 & 0.71 & -0.74 \\
0.71 & 0.64 & 0.64 & -0.66 \\
0.71 & 0.64 & 0.64 & -0.66 \\
-0.74 & -0.66 & -0.66 & 0.68
\end{pmatrix}
\end{eqnarray}
where $\circ$ is the element-wise product and $\mathcal{N}=\operatorname{Tr}(\rho_a\circ G_{\textnormal{herald}})$ is the normalization factor. Here the diagonal elements of $G_{\textnormal{herald}}$ indicate the probability of a heralding event (for unity detection efficiency), which depends slightly on the atomic state. The persisting minus-signs in the fourth row and column are a characteristic of the controlled-Z gate.

If no heralding is applied and the atomic state is accepted for any final photonic mode, the photonic state has to be traced out:
\begin{eqnarray}
\rho_{a,\textnormal{out}}^{\textnormal{total}} &=& \operatorname{Tr}_p \sum_{i,j=0}^3 a_{i,j}\ket{a_i}\bra{a_j}\otimes\ket{p_{i}}\bra{p_{j}}
=\nonumber\\
&=& \sum_{i,j=0}^3 a_{i,j}\ket{a_i}\bra{a_j}\bra{p_{j}}{p_{i}}\rangle=\nonumber\\
&=:& \rho_a\circ G_{\textnormal{total}}=\nonumber\\\label{eq:operator_noherald}
&=& \rho_a\circ
\begin{pmatrix}
1 & 0.90 & 0.90 & -0.72 \\ 
0.90 & 1 & 0.64 & -0.62 \\ 
0.90 & 0.64 & 1 & -0.62 \\ 
-0.72 & -0.62 & -0.62 & 1
\end{pmatrix}
\end{eqnarray}
In this case, only off-diagonal elements are reduced, as a result of the decoherence from the photon leaking information into the environment.

\subsection{Fidelity of the two-atom entangling operation}
The initial atomic state for our entangling operation is $\rho_{a,\textnormal{in}}=\ket{a_\textnormal{in}}\bra{a_\textnormal{in}}$ with $\ket{a_\textnormal{in}} = \frac12\left(\ket{\up\up}-\ket{\up\down}-\ket{\down\up}+\ket{\down\down}\right)$. Reflecting a single photon without heralding applies the operator (\ref{eq:operator_noherald}) $\rho_{a,\textnormal{out}}^{\textnormal{total}}=\rho_{a,\textnormal{in}}\circ G_{\textnormal{total}}$. Without further imperfections the entangled state fidelity would thus be
\begin{equation}
\label{eq:fidelity_noherald}
F = \bra{\Phi^-}\rho_{a,\textnormal{out}}^{\textnormal{total}}\ket{\Phi^-} = 79.2\%.
\end{equation}

When the reflection of a single photon is heralded, the operator $G_{\textnormal{herald}}$ (\ref{eq:operator_herald}) is applied, and the resulting density matrix must be normalized $\rho_{a,\textnormal{out}}^{\textnormal{herald}}=\rho_{a,\textnormal{in}}\circ G_{\textnormal{herald}} / \operatorname{Tr}(\rho_{a,\textnormal{in}}\circ G_{\textnormal{herald}})$. In this case, the fidelity
\begin{equation}
\label{eq:fidelity_herald}
F = \bra{\Phi^-}\rho_{a,\textnormal{out}}^{\textnormal{herald}}\ket{\Phi^-} = 99.7\%
\end{equation}
is close to unity, as the only remaining imperfection is the slightly differing reflection probability for each number of coupling atoms.

In the actual experiment we employed weak coherent pulses with a mean photon number $\bar{n} = 0.13$, which are superpositions of Fock-states $\ket{n}$ with a Poissonian distribution $P(n)=\bar{n}^n/n!\cdot e^{-\bar{n}}$. To model the evolution of the atomic state, we averaged the density matrix over all possible combinations of cavity-coupled and non-coupled photons as well as detector dark counts. Each case was weighted with the corresponding posterior probability to receive one heralding click in the photon detector. In case of a dark count without impinging photon the atoms remain in the input state having $25\%$ overlap with the anticipated entangled state. If one photon is reflected, the operations (\ref{eq:operator_herald}) or (\ref{eq:operator_noherald}) apply, depending on whether the photon gets lost in the process. If several photons are reflected, the operators are applied multiple times, leading to an increased decoherence.

Considering our total photon detection efficiency of $46\%$ (including $55\%$ efficiency of the single-photon detector), a dark-count rate of $0.2\%$ per pulse and no additional imperfections apart from the cavity, we obtain an entangled state fidelity of $91.1\%$ for an impinging weak coherent pulse with $\bar{n}=0.13$. In the following we include the known significant sources of errors, which are listed in Table I, in our simulation. The main sources of error are imperfect mode matching of the incoming photons to the cavity ($92\%$ overlap) and imperfect state detection of the atoms ($3\%$ of all experiments). Multiphoton contributions are present in $8\%$ of all pulses that lead to a heralding event. The error introduced by loosing a photon in the cavity is determined by Eq.~(\ref{eq:operator_herald}) and in case of dark counts by Eq.~(\ref{eq:operator_noherald}). We estimate the photon polarization to be accurate to $1\%$. The atomic state preparation was wrong in around $1\%$ of all cases and their decoherence during $14\unit{\micro{}s}$ between preparation and readout is around $1\%$ as well. With these errors the simulation yields an entangled state fidelity of $77\%$, which is fits well with the measured value of $76(2)\%$. By switching each source of error on and off in the simulation, we determined its influence onto the fidelity. The results are listed in Table I in the main text.

\subsection{Expected entangled-state fidelity employing a single-photon~source}
All presented experiments were performed with weak coherent pulses. This lowers the overall efficiency due to necessary post-selection on single-photon detections and leads to fidelity reductions caused by the admixture of higher photon-number states and dark count events. As this limitation is not intrinsic to the gate scheme and could be overcome in the foreseeable future, it is worthwhile to estimate the fidelity of the atom-atom gate that could be achieved by employing an ideal external single-photon source.

Replacing the weak coherent pulses by true single photons, the density matrix calculation with the known sources of fidelity reductions (Table~\ref{tab:errors}) can be applied in the same way, only without the additional summing over different input photon numbers. With single photons and the heralding still applied, the fidelity would then increase from $76\%$ to $82\%$ and the efficiency would jump from $4.2\%$ to $32\%$. The latter is the reflection, coupling and detection efficiency of a photon. The main improvement in fidelity would be through the elimination of multiphoton contributions. Fidelity reductions from photon loss in the cavity would play only a negligible role [$0.3\%$, Eq.~(\ref{eq:fidelity_herald})] because the heralding ensures that the one impinging photon must have survived. The remaining infidelities are mostly due to the imperfect optical mode matching and the final state detection of the atoms.

With a single-photon source, the heralding could be dropped altogether, making the gate fully deterministic with an efficiency of $100\%$. In this case our experimental imperfections would lead to a fidelity of $68\%$ according to the simulation. Thus without heralding the efficiency would increase by a factor of three while the fidelity would only drop by 14\%, because there is a considerable probability that the gate worked perfectly even if the inefficient photon detector didn't yield a click. The dominating source of errors would be the photon loss within the cavity [Eq.~(\ref{eq:fidelity_noherald})]. Additional errors are predominantly the optical mode matching and the imperfect state detection. The total fidelity of $68\%$ is lower than the measured fidelity for heralded weak coherent photons, but it is still significantly above the classical threshold of $50\%$.

\section{EVALUATION OF THE TRUTH TABLE MEASUREMENTS}\label{sec:truthtablemethods}
\begin{figure}[b]
\centering
\hypertarget{fig:parity_cnot}{}
\includegraphics[width=8cm]{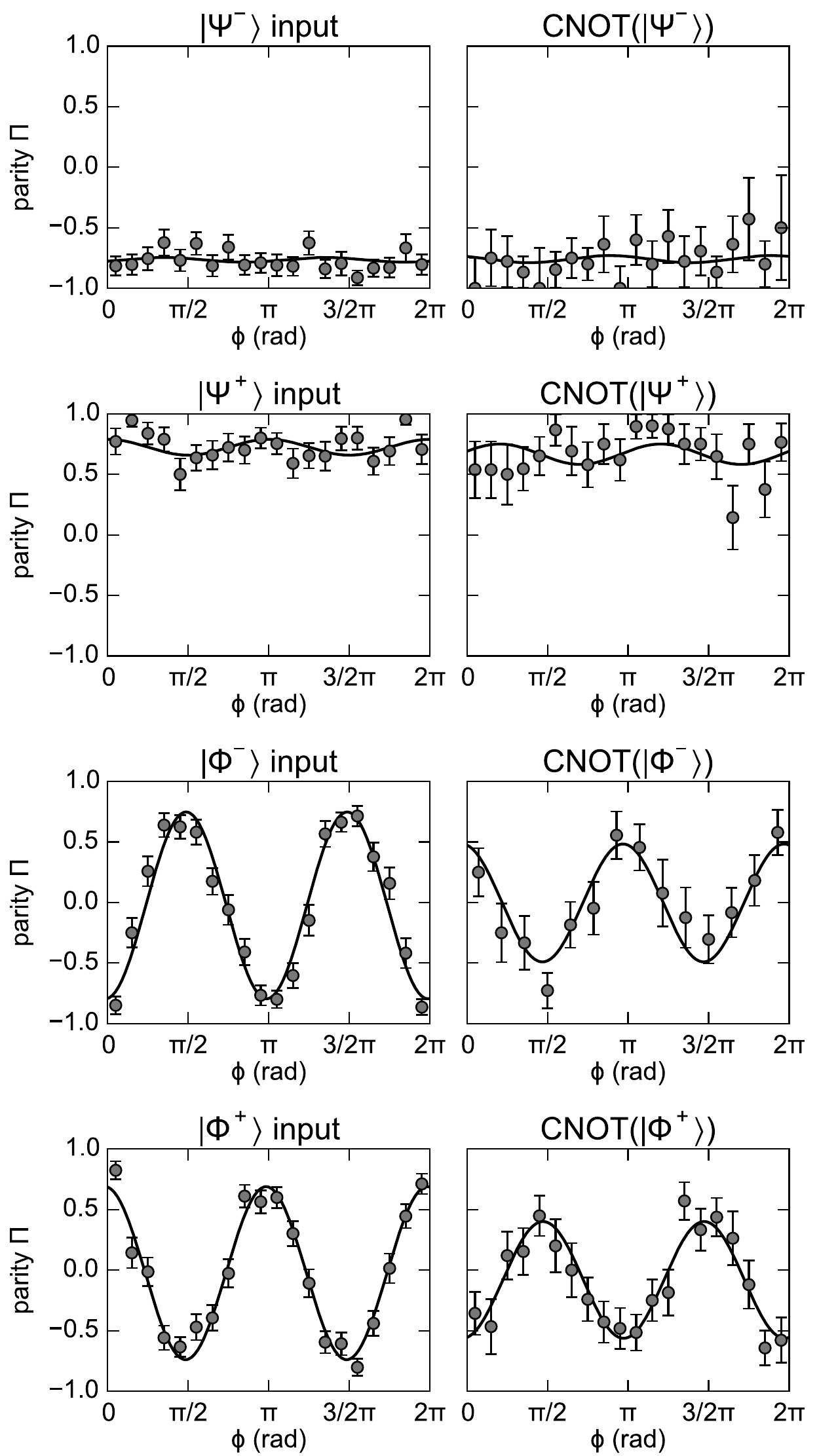}
\caption{\label{fig:parity_cnot}%
Parity signals of input and output states of the CNOT gate. The left column shows $\Pi(\phi)$ for all four prepared Bell states used as input states for the gate. The right column shows the respective parity signals after the gate. While the states $\ket{\Psi^\pm}$ remain basically unchanged by the gate operation, one can observe the swap $\ket{\Phi^\pm}{\rightarrow}\ket{\Phi^\mp}$ as a $\pi/2$ phase flip in the respective parity oscillation signal. Error bars are statistical standard errors of each data point. The solid lines are fitted sine curves with free offset, amplitude and phase, which are used to compute the fidelities.}
\end{figure}

In our gate characterization measurement for the controlled-NOT (CNOT) operation, we use the Bell basis states $\ket{\Psi^\pm}$ and $\ket{\Phi^\pm}$ as input, prepared as detailed in the preceding section. The gate is expected to produce Bell states as output. The fidelities of our measured states with the ideal ones, depicted in Fig.~\figref{fig:truthtable}, are determined from average state populations $P_{\up\up}$, $P_{\up\down}+P_{\down\up}$ and $P_{\down\down}$ and parity oscillations $\Pi(\phi)$. These parity oscillations are recorded after a common $\pi/2$ rotation on both atomic qubits with a rotation axis determined by the relative phase $\phi$ to previous rotations, which we imprint on the Raman laser beams. A population measurement after such an analysis pulse reveals coherence properties of the two-qubit state. This can be seen from the analytical expression of the parity as a function of $\phi$:
\begin{equation}
\begin{array}{rcl}
 \Pi(\phi)&=&\tilde P_{\up\up}(\phi)+\tilde P_{\down\down}(\phi)-\bigl(\tilde P_{\up\down}(\phi){+}\tilde P_{\down\up}(\phi)\bigr)\\[1ex]
 &=&2\mathop{\mathrm{Re}}(\rho_{\up\down,\down\up})+2\mathop{\mathrm{Re}}(\rho_{\up\up,\down\down})\cos(2\phi)\\
&&\hphantom{2\mathop{\mathrm{Re}}(\rho_{\up\down,\down\up})}+2\mathop{\mathrm{Im}}(\rho_{\up\up,\down\down})\sin(2\phi)
\end{array}
\end{equation}
The information about the respective populations and coherences is sufficient to determine the fidelities with each of the four Bell states according to Eq.~(\ref{eq:bellfideliy}):
\begin{equation}
\label{eq:bellfideliysupp}
\begin{array}{rcl}
F(\ket{\Psi^-})&=&\textstyle\frac12(P_{\up\down}+P_{\down\up})-\mathop{\mathrm{Re}}(\rho_{\up\down,\down\up})\\[1ex]
F(\ket{\Psi^+})&=&\textstyle\frac12(P_{\up\down}+P_{\down\up})+\mathop{\mathrm{Re}}(\rho_{\up\down,\down\up})\\[1ex]
F(\ket{\Phi^-})&=&\textstyle\frac12(P_{\up\up}+P_{\down\down})-\mathop{\mathrm{Re}}(\rho_{\up\up,\down\down})\\[1ex]
F(\ket{\Phi^+})&=&\textstyle\frac12(P_{\up\up}+P_{\down\down})+\mathop{\mathrm{Re}}(\rho_{\up\up,\down\down}).
\end{array}%
\end{equation}%

\begin{table}
\caption{\label{tab:cnotdata}%
State populations and fidelities.
The experimentally measured populations $P_{\up\up}$, $P_{\up\down}{+}P_{\down\up}$ and $P_{\down\down}$ before and after the gate operation are listed. Furthermore, the table contains the offset $\overline{\Pi}:=2\mathop{\mathrm{Re}}(\rho_{\up\down,\down\up})$ and the cosine amplitude $\Delta\Pi:=2\mathop{\mathrm{Re}}(\rho_{\up\up,\down\down})$ of all four parity signals from the fits in Fig.~\protect\figref{fig:parity_cnot}. From this data, the respective fidelities $F$ can be inferred.}
\begin{ruledtabular}%
\begin{tabular}{ccccccc}%
state & $P_{\up\up}$ & $P_{\up\down}{+}P_{\down\up}$ & $P_{\down\down}$ & $\overline{\Pi}$ & $\Delta\Pi$ & $F$ \\
\hline
\hphantom{CNOT(}$\ket{\Psi^-}$\hphantom{)} & \03\% &  91\% & \06\% &  $-0.76$ &  $-0.01$ & 84(1)\% \\
          CNOT($\ket{\Psi^-}$)             & \05\% &  86\% &  10\% &  $-0.76$ & $\m0.02$ & 81(3)\% \\
\hline
\hphantom{CNOT(}$\ket{\Psi^+}$\hphantom{)} & \07\% &  84\% & \09\% & $\m0.72$ & $\m0.06$ & 78(2)\% \\
          CNOT($\ket{\Psi^+}$)             & \03\% &  91\% & \06\% & $\m0.67$ & $\m0.02$ & 79(3)\% \\
\hline
\hphantom{CNOT(}$\ket{\Phi^-}$\hphantom{)} &  29\% &  15\% &  56\% &  $-0.02$ &  $-0.77$ & 81(2)\% \\
          CNOT($\ket{\Phi^-}$)             &  23\% &  12\% &  65\% & $\m0.00$ & $\m0.48$ & 68(4)\% \\
\hline
\hphantom{CNOT(}$\ket{\Phi^+}$\hphantom{)} &  45\% &  15\% &  40\% &  $-0.03$ & $\m0.71$ & 78(2)\% \\
          CNOT($\ket{\Phi^+}$)             &  44\% &  10\% &  46\% &  $-0.08$ &  $-0.47$ & 69(3)\%
\end{tabular}%
\end{ruledtabular}%
\end{table}
Figure~\figref{fig:parity_cnot} shows the recorded parity oscillations for the two-atom states before and after the gate, and Table~\ref{tab:cnotdata} lists the state populations and resulting fidelities that produce the bar plots depicted in Fig.~\figref{fig:truthtable}.

%\bibliographystyle{./bibstyle}
%\bibliography{./references}

\end{document}